\documentstyle[prl,aps,twocolumn]{revtex}
\begin{document}
\draft

\title{On the Penrose Inequality for general horizons}
\author{Edward Malec}
\address{Instytut Fizyki, Uniwersytet Jagiello\'nski,
Reymonta 4, P- 30-059 Krak\'ow, Poland}
\author{Marc Mars}
\address{Max-Planck-Institut f\"ur Gravitationsphysik,
Am M\"uhlenberg 1, D-14 476 Golm, Germany}
\author{Walter Simon}
\address{Institut f\"ur theoretische Physik der Universit\"at Wien,
Boltzmanngasse 5, A-1090 Wien, Austria}

\maketitle

\begin{abstract}

For asymptotically flat initial data of Einstein's equations satisfying an energy condition, we show 
that the Penrose inequality holds between the ADM mass and the area of an outermost apparent 
horizon, if the data are restricted suitably. We prove this by generalizing Geroch's proof 
of monotonicity of the Hawking mass under a smooth inverse mean curvature
flow, for data with non-negative Ricci scalar. Unlike Geroch  we need not
confine ourselves to minimal surfaces as horizons. Modulo smoothness issues we also show that our 
restrictions on the data can locally be fulfilled by a suitable choice of the initial surface in a 
given spacetime.

\end{abstract}


\date{today}

An important issue in General Relativity is the ``cosmic censorship conjecture'' which reads, 
roughly speaking, that singularities (which necessarily develop in gravitational collapse 
in the future of apparent horizons) are always separated from the outside world by event 
horizons. 
%In attempts of  proving or disproving this conjecture, 
Penrose gave a heuristic argument which showed that, for collapsing shells of particles with 
zero rest mass, cosmic censorship would imply the inequality

\begin{equation}
\label{pi}
M \ge \sqrt{\frac{A}{16\pi}}
\end{equation}
between the total mass (the ADM-mass) $M$ of the spacetime and the area $A$ of an apparent horizon 
${\cal H}$ \cite{RP}.
While for such shells inequality (\ref{pi}) was recently proven by Gibbons
\cite{GG}, there has remained 
the challenge of proving (independently of cosmic censorship) the Penrose inequality (PI) (\ref{pi}) in the 
``general case'' characterized  below.

We consider a spacetime $({\cal M},{}^4g_{\mu\nu})$ and a corresponding initial data set 
$({\cal N},g_{ij},k_{ij})$, i.e. a smooth 3-manifold ${\cal N}$ (which can be embedded in ${\cal M}$), 
a positive definite metric $g_{ij}$ and a symmetric tensor $k_{ij}$ (the second fundamental form of the embedding). 
To be compatible with Einstein's equations on ${\cal M}$, these quantities satisfy the constraints
\begin{eqnarray}
\label{con1}
D_i \left( k^i_l - k \delta^i_l \right) & = & - 8 \pi j_l, \\
\label{con2}
R - k_{ij}k^{ij} + k^2 & = & 16\pi\rho.
\end{eqnarray} 
where $D_{i}$ is the covariant derivative and $R$ the Ricci scalar on ${\cal N}$, $k = g^{ij}k_{ij}$, 
and $\rho$ and $j_{i}$ are the energy density and the matter current, respectively. We take the data to be 
asymptotically flat and to satisfy the dominant energy condition (which implies that $\rho \ge |j|$). 
Furthermore, we assume that the boundary of ${\cal N}$ (if non-empty) is a ``future apparent horizon'' 
${\cal H}$  (called a ``horizon'' from now on) which is a 2-surface defined by the property that all outgoing 
future directed null geodesics (in ${\cal M}$) orthogonal to ${\cal H}$ have vanishing divergence, i.e.  
$\theta_{+} = 0$, and that the divergence of the outgoing past null geodesics is non-negative, i.e. $\theta_{-} \ge 0$. 
(Past apparent horizons are defined similarly, and satisfy analogous theorems). 
If $\theta_{+} = \theta_{-} = 0$ on ${\cal H}$, the horizon 
is an extremal surface of $({\cal N}, g_{ij})$ and the outermost extremal surface in an asymptotically flat space 
must be minimal.

As an idea for proving the positive mass theorem (PMT) Geroch considered an asymptotically flat 
Riemannian manifold  $({\cal N}, g_{ij})$ with non-negative Ricci scalar $R$ (which naturally arises 
from initial data sets described above by restricting $k_{ij}$ suitably) and
assumed that on $({\cal N},g_{ij})$ there is a smooth ``inverse mean curvature flow''
(IMCF). This means that one can write the metric $g_{ij}$ as

\begin{equation}
\label{met}
ds^2 = \phi^2 dr^2 + q_{AB}dx^A dx^B
\end{equation}
(where $A,B = 2,3$), with smooth fields $\phi$ and $q_{AB}$,  and with $p \phi = 1$ where $p$ is the mean 
curvature of the surfaces $r= \mbox{const}$. (This parametrization yields $dA/dr = A$, i.e. the area 
$A$ of the surfaces increases exponentially). Assuming also that these surfaces have spherical topology, 
Geroch showed that the mass functional 
\begin{equation}
\label{mg}
M_{G}({\cal S}) = \frac{\sqrt{A}}{64\pi^{\frac{3}{2}}} 
\left (16\pi - \int_{\cal S} p^2 dS \right)
\end{equation}
is monotonic under a smooth IMCF, i.e. $dM_{G}/dr \ge 0$ \cite{RG}. In some cases (in particular, when $k_{ij}=0$) 
$M_{G}$ is a special case of the functional
\begin{equation}
\label{mh}
M_{H}({\cal S}) = \frac{\sqrt{A}}{64\pi^{\frac{3}{2}}} 
\left (16\pi - \int_{\cal S} \theta_{+}\theta_{-} dS \right) 
\end{equation}
introduced earlier by Hawking \cite{SH}.
Recently, Huisken and Ilmanen proved that monotonicity of $M_G$
also holds without the smoothness assumption on the flow 
\cite{HI1,HI2}. Since $M_{G}$ tends to $M$ at spatial infinity $I^0$, flowing by IMC out of a point proves the 
positive mass theorem ($M \ge 0$) on manifolds without minimal surfaces (for which  $p = 0$). On the other hand, since 
$M_{G}$ equals $\sqrt{A/16\pi}$ on a minimal surface, flowing from the outermost such surface to infinity proves 
the special case of the PI (\ref{pi}) for which horizons coincide with minimal surfaces \cite{HI1}-\cite{JW}.

In the present Letter we show that the Hawking mass (\ref{mh}) 
is also monotonic under a smooth IMCF provided that certain 
rather simple supplementary conditions are satisfied. To motivate the latter and to 
compare them with previous work, it is useful to formulate first, as a
conjecture, the sharpened version 

\begin{equation}
\label{gpi}
M^2 \ge \frac{A}{16\pi} + P^2.
\end{equation}

of the PI (c.f. \cite{GH} for a careful formulation). This inequality seems natural in view of the PMT $M \ge |P|$ 
for the ADM-momentum $P$ \cite{SY1,EW} (which also holds in the presence of apparent horizons \cite{GHHP,MH}). 
Note that the anticipated result $(\ref{gpi})$ only involves the area $A$  and the norm $M^2 - P^2$ of the 
ADM 4-vector $P^{\mu}$ which are {\it spacetime quantities}. On the other hand, we wish to show monotonicity of 
(\ref{mh}) for a class of {\it data} $({\cal N},g_{ij},k_{ij})$ as general as possible. 
%In short, we wish to get a result about 
%{\it spacetimes} by using a non-unique description in terms of their {\it data}. 
For this purpose we can now pursue either an ``invariant'' approach in which an arbitrary hypersurface ${\cal N}$ 
connecting ${\cal H}$ with $I^{0}$ is admitted, or a ``gauge'' approach for which ${\cal N}$ is fixed suitably. 
While in the former setting we obtain monotonicity of $M_H$ only for a restricted class of data,
this monotonicity then directly implies the PI since (like $M_G$) $M_H$ still tends to $\sqrt{A/16\pi}$ on the 
horizon and to $M$ at $I^0$. On the other hand, in the "gauge approach" we can implement restrictions on the data
by determining ${\cal N}$ via initial conditions on ${\cal H}$ and via a suitable {\it local} propagation law. However, 
here we cannot guarantee that ${\cal N}$ really ends up at $I^0$  rather than becoming hyperboloidal and 
approaching null infinity.

As to the ``invariant'' approach, Jang has presented a generalization of the functional 
(\ref{mg}) which is, for {\it any} data $({\cal N}, g_{ij}, k_{ij})$,  monotonic when propagated 
outwards with a flow determined by a certain quasilinear partial differential equation \cite{PJ1}. 
If existence of solutions for this equation could be proven, then 
it would follow that $M \ge |P|$, and $M = 0$ would imply that 
$({\cal N}, g_{ij}, k_{ij})$ are data for flat space. 
Unfortunately, Jang's equation is too complicated to be tractable at present.  Moreover, as to possible 
extensions towards a PI, the boundary terms at the horizon seem to be difficult to analyze in this method.

On the other hand, the ``gauge'' approach to proving the PI has been put forward by Frauendiener 
\cite{JF}. He suggests to construct a foliation ${\cal S}(r)$ and a hypersurface 
${\cal N}$ by starting from ${\cal H}$ and moving outwards with the ``inverse mean curvature vector'' 
(assuming this is spacelike),

\begin{equation}
\label{mcv}
J^{\mu} = \frac {1}{\theta_{+}}l^{\mu}_{+} + \frac {1}{\theta_{-}}l^{\mu}_{-}
= \frac{\sqrt{2}}{\theta_{+}\theta_{-}} \left(p m^{\mu} - q n^{\mu} \right),
\end{equation}
where $l^{\mu}_{+}$ and $l^{\mu}_{-}$ are, respectively,
tangent to the future and past outgoing null geodesics emanating orthogonally from a level surface
${\cal S}(r)$, with $l^{\mu}_{+} l_{- \mu} = 1$, $\sqrt{2} m^{\mu} = l^{\mu}_{+} + l^{\mu}_{-}$, 
$\sqrt{2} n^{\mu} = l^{\mu}_{+} - l^{\mu}_{-}$, $2p = \theta_{+} + \theta_{-}$ and 
$2q = \theta_{+} - \theta_{-}$. While $J^{\mu}$ is uniquely defined (by the first equation (\ref{mcv})), $m^{\mu}$, 
$n^{\mu}$, $p$ and $q$ depend on the scaling $l^{\mu}_{+} \rightarrow \lambda l^{\mu}_{+}$, 
$l^{\mu}_{-} \rightarrow \lambda^{-1} l^{\mu}_{-}$ (with some function
$\lambda$ on ${\cal S}(r)$). By choosing $\lambda$ suitably we can achieve that $m^{\mu}$ and $n^{\mu}$ are, 
respectively, parallel and orthogonal to any spacelike hypersurface ${\cal N}$. In particular, when $J^{\mu}$ 
is required to be tangent to ${\cal N}$, we have $q \equiv 0$ there.  This
so-called ``polar hypersurface condition''  \cite{BP} implies $R \ge 0$, 
and it also implies that ${\cal N}$ can reach only those horizons (where $\theta_{+} = 0$) which also satisfy $p = 0$
and are thus minimal surfaces. 
Therefore Frauendiener's approach in essence boils down to the situation considered by Geroch already {\cite{RG}. 
An advantage of Frauendiener's  4-dimensional reformulation is that he can set out from the ``weak energy condition'' 
to {\it obtain} $R \ge 0$ instead of ${\it imposing}$ the latter condition. However, as mentioned before, it is not 
clear whether the constructed hypersurface really reaches $I^0$. 

In the theorem below, we prove the ``weak form'' (\ref{pi}) of the PI for ``general'' future apparent horizons, 
within the "invariant aproach" and for a restricted class of data. We will then show that, modulo smoothness 
issues, these restricions can be omitted within a "gauge" approach where, in fact, a large family of
hypersurfaces is admitted.

The following definitions apply to a smooth (but otherwise arbitrary) foliation of
${\cal N}$ by 2-surfaces ${\cal S}(r)$. Let $m^i$ denote the unit normal to ${\cal S}(r)$,  $\phi$ the 
``lapse'' (c.f. (\ref{met})), $q_{ij} = g_{ij} - m_i m_j$ the induced metric, $p_{ij}$ the second fundamental 
form of ${\cal S}$ in ${\cal N}$, $t_{ij}$ its trace-free part (i.e.  $t_{ij} = p_{ij} - (p/2) q_{ij}$) and 
$p$ its trace (the mean curvature). We will also employ the following decomposition of $k_{ij}$ w.r. to ${\cal S}(r)$, 

\begin{equation}
\label{kdec}
k_{ij} = z m_i m_j +  m_{i}s_{j} + m_{j}s_{i} + q_i^k q_j^l x_{kl} + \frac{1}{2}q q_{ij},
\end{equation}
where $z = k_{ij}m^{i}m^{j}$, $s_i = q_i^j k_{jl}m^{l}$, $q = k_{ij}q^{ij}$
 and $x_{ij} = q^l_i q^n_j k_{ln} - (1/2)q q_{ij}$.
From $p$ and $q$ we can define
$\theta_+$ and $\theta_-$ using the expressions following (\ref{mcv}). \\ \\
{\it Theorem}. Let $({\cal N},g_{ij},k_{ij})$ be a smooth, asymptotically flat initial data set 
for Einstein's equations with an outermost future apparent horizon ${\cal H}$ 
(i.e. $\theta_{+} = 0$ and $\theta_{-} \ge 0$  on  ${\cal H}$) of spherical
topology, which satisfies the constraints (\ref{con1}),(\ref{con2}), the dominant energy condition,
and the following additional restrictions
\begin{enumerate}
\item On $({\cal N},g_{ij})$ there exists a smooth inverse mean curvature flow.
\item The divergence $\theta_{-}$ (taken with respect to the level sets of
the IMCF) is positive outside ${\cal H}$.
\item On each level set of the IMCF, at least one of (a) or (b) holds.
\begin{enumerate}
\item $q/p$ is constant.
\item $q^{ij}D_{i}s_{j} = 0$, and $\theta_{+} \ge 0$.
\end{enumerate}
\end{enumerate}

Then the PI (\ref{pi}) holds. \\ \\
{\it Proof.} We show monotonicity of the Hawking mass functional (\ref{mh}). 
Noting that $\theta_{+}\theta_{-} = p^2 - q^2$, we compute the derivatives of $p$ and $q$ 
in the direction of $m^i$ (the IMCF is not required for this step). 
Using (\ref{kdec}), the 3- and the 2-dimensional constraints and the fact that 
$\phi m^i D_i m_k = - q_k^j D_{j} \phi$ in the coordinates (\ref{met}) gives

\begin{eqnarray}
2 m^i D_i p & = & - 2 \frac{{}^{2}\Delta \phi}{\phi}
- 16\pi\rho - 2 s_{i}s^{i} - \nonumber \\
&&x_{ij}x^{ij} + \frac{1}{2}q^2 + 2 z q + {}^{2}R - \frac{3}{2} p^2 -  t_{ij}t^{ij}, \nonumber
\\
m^i D_i q  &=&  8\pi j_i m^i - x_{ij}t^{ij} + p(z - \frac{1}{2}q) + \nonumber \\
&&q^{ij}D_{i}s_{j} + 
2 \frac{s^i D_i \phi}{\phi},
\end{eqnarray}
where ${}^{2}R$ and ${}^{2}\Delta$ are the Ricci scalar and the Laplacian
with respect to $q_{ij}$.
Next we use that $m^i D_{i}\sqrt{g} = p \sqrt{g}$, restrict ourselves to an IMCF
(i.e. $ \phi p = 1$) and remove the ${}^{2}\Delta$-term via integration by parts.
We obtain, for the Lie derivative ${\cal L}_{\phi m^i}$ of $M_{H}$,

\begin{eqnarray}
{\cal L}_{\phi m^i} M_{H}({\cal S}) & = & \frac{\sqrt{A}}{64\pi^{3\over 2} }
\int_{\cal S} \Biggl[ 16\pi \left( \rho +   
{q\over p} j_i m^i \right) + \nonumber \\
&& + \left( x_{ij}x^{ij} - 2{q \over p}x_{ij}t^{ij} + t_{ij}t^{ij} \right)
+  \nonumber \\
&& + 2 \left( s_{i}s^{i} - 2\frac{q}{p} s^i \frac{D_i p}{p} + 
\frac{q^{ij} (D_i p)(D_j p)}{p^2} \right) + 
\nonumber\\
&& + 2 \frac{q}{p} q^{ij}D_i s_j \Biggr] dS.
\label{Lm}
\end{eqnarray}

We first show that $|q/p| \le 1$ which is obvious when conditions 2. and 3.(b) hold.
 If in turn we assume conditions
2. and 3.(a), we can write $\theta_{+} =\gamma \theta_{-}$ where $\gamma = (1+q/p)/(1-q/p)$ is constant on each 
${\cal S}$.
Since $\theta_{+} > 0$ at some point exterior to ${\cal H}$, the same
is true for $\gamma$. If $\gamma$ were to vanish
somewhere in the exterior, it must vanish on the whole leaf ${\cal S}$,
in which case ${\cal H}$ would not be the outermost horizon. 
Therefore, $\gamma > 0$ and $\theta_{+} > 0$ in the exterior, 
which implies our claim.

Now the dominant energy condition together with $|q/p| \le 1$ implies that the first term (in parentheses) 
on the r.h.  side of (\ref{Lm}) is non-negative. Next, again due to $|q/p| \le 1$, the second and the 
third terms are
positive quadratic forms. Finally, the last term vanishes obviously by condition 3.(b),
by partial integration on ${\cal S}(r)$ due to 3.(a). Hence  ${\cal L}_{\phi m^i} M_{H} \ge 0$, and integrating 
(\ref{Lm}) between the horizon and infinity finishes the proof.\\

We remark that conditions 2. and 3. could be substituted by any weaker condition which still makes the r.h. 
side of (\ref{Lm}) non-negative (in particular we could just demand this property). The reason for having 
selected  2. and 3. is that these conditions seem to leave enough freedom such that, in a given spacetime, 
they could always be satisfied by a suitable choice of the hypersurface. Moreover,
they seem simple enough such that this conjecture could be proven.
In fact, consider a  spacetime ${\cal M}$ and an arbitrary (smooth) function $F(r)$  with $F(0)=-1$ 
and $|F(r)| < 1$ for $r > 0$. 
Assuming that "everything" is smooth, we can then find a hypersurface ${\cal N}$ 
such that the data induced on ${\cal N}$ satisfy 1., 2. and 3.(a), where 
$r = const$ are the level sets of the IMCF, with ${\cal H}$ located at $r = 0$, and with $q/p = F(r)$. 
To see this on a heuristic basis, assume we have found a "piece" of such a hypersurface, bounded by a level set 
${\cal S}_0$ (given by $r = r_0$) of the IMCF. 
Rescaling the null vectors $l^{\mu}_{-}$ and $l^{\mu}_{+}$ emanating from ${\cal S}_0$  like 
$l^{\prime \mu}_{+} = \lambda l^{\mu}_{+}$ and $l^{\prime \mu}_{-} = \lambda^{-1} l^{\mu}_{-}$, with 
$\lambda^2 = \theta_{-} \theta_{+}^{-1}(1 + F(r))(1 - F(r))^{-1}$ (and keeping  $\lambda$ constant along 
$l^{\mu}_{+}$ and $l^{\mu}_{-}$) the spacelike vector field 
$\sqrt{2} m^{\prime \mu} = l^{\prime \mu}_{+} + l^{\prime \mu}_{-}$ then determines 
the direction in which ${\cal N}$ has to be continued across ${\cal S}_0$. 
Since $F(r)$ can be chosen arbitrarily, it is also plausible that a suitable
choice would make the surface ${\cal N}$ reach $I^0$. 

We now turn to the issue of removing condition 1., both for "arbitrary"  and for "selected" hypersurfaces. 
Recall that, in the "time-symmetric" context, Huisken and Ilmanen have defined a generalized (``weak'') 
IMCF flow \cite {HI1,HI2}. The latter ``jumps'' at those (countably many) parameter values for which the set 
swept out by the flow can be enclosed by another one of the same area but with larger volume. Due to the 
requirement on the area, the additional segments of the new hull necessarily consist of minimal surfaces, 
i.e. $p = 0$. Such segments do not contribute to $M_G$, whence the latter 
{\it increases} at each jump. In the more general situation considered here, these ``jumps'' seem still 
appropriate since they go well with IMCF. However, they will not directly preserve monotonicity of
$M_{H}$ in general. Though probably very difficult, this problem
could be addressed in several ways. First, one could take a pure ``initial data'' perspective, in which the
constraints have to be solved at the same time as the flow. In this case, the value of $q$ after the jump is
not given a priori but still needs to be obtained. Thus, it is conceivable that general enough initial data exist
for which $M_H$ is still monotonic at the jumps. Alternatively, one could take a spacetime
point of view. It may be possible to generalize the procedure for defining a hypersurface sketched above for 
the smooth case, to obtain a weak flow such that 2. and 3. are satisfied and $M_H$ is monotonic. Instead, it may 
turn out to be necessary to allow for more general flows and let the two-surfaces ``jump in spacetime'', by which we 
mean that they are not a priori restricted to {\it any 3-surface whatsoever}. In any case, it seems reasonable to 
{\it conjecture} that the theorem above holds if ``initial data set'' is replaced by ``spacetime'', and if all 
{\it conditions 1. - 3. are removed}. 

Some further comments are in order.

Equation (\ref{Lm}) can be written in integral form as $M_H({\cal S}_2)=M_H({\cal S}_1) + M_V$. 
Here $M_V = \int_{V}B d\varrho dS/16\pi
$, $B$ stands for the integrand in (\ref{Lm}), $\varrho =\sqrt{A/4\pi }$ is an areal radius
coordinate, ${\cal S}_1$ and ${\cal S}_2$ are two leaves of the foliation and
$V({\cal S})$ is the volume of the annulus enclosed by ${\cal S}_1$ and ${\cal S}_2$. 
$M_V$ is nonnegative under the assumptions stated in the theorem. When ${\cal S}_1$ reduces to a point, 
$ M_V$ can be regarded as a volume representation of the Hawking mass $M_H({\cal S}_2)$.  
If the internal boundary ${\cal S}_1$ lies outside ${\cal H}$, we obtain $M \ge \sqrt{A/16\pi} +
M_{V}$ (where $A$ is still the area of ${\cal H})$.
Notice that outside ${\cal H}$ the quantity $M_V$ can be 
useful to give a bound from above to the energy norm of matter fields; this may become relevant in the 
investigation of the Cauchy problem. 
 
As to known special cases of our formula (\ref{Lm}), it reduces for $k_{ij} = 0$ 
%(i.e. for $q = s_i= x_{ij} = 0$) 
to the expression obtained by Geroch \cite{RG} and for $q = 0$ (upon translating to spinor formalism) to 
Frauendiener's expression (equ. (9) of \cite{JF}). In the spherically symmetric case (i.e. when $g_{ij}$ and 
$k_{ij}$ are invariant under rotations) the PI was proven by Malec and O'Murchadha \cite{MM} and by Hayward \cite{SeH1}. 
This result can also be recovered from the theorem above, since the level surfaces of the IMCF are
metric spheres in this case, and each of the conditions in 3. is obviously satisfied. 

Compared with Jang's ideas for proving the PMT for general data, we need to restrict our data strongly, but we 
require only the knowledge of the IMCF rather than proving existence for an involved quasilinear
PDE. We note that Jang sets out from a mass functional which is different from (\ref{mh}), even after imposing our 
gauge condition 3. 
Therefore, imposing the latter in Jang's computation will not yield the present results.

Our approach might also be useful to prove the PI involving the Bondi mass rather than the ADM one. 
Under suitable technical assumptions, this version of the PI has been obtained by Ludvigsen and Vickers \cite{LV} 
and by Bergqvist \cite{GB} by showing that certain mass functionals (different from (\ref{mh})) are monotonic 
along {\it null hypersurfaces}, and that they have  the appropriate values at the horizon and at null infinity. 
The same technique has been applied to the Hawking mass (\ref{mh}) by Hayward  \cite{SeH2} who obtains a
pair of monotonicity formulas for $M_{H}$ along the two null directions. We note that
by just taking linear combinations of Hayward's expressions one does {\it not} obtain the
Lie derivative of $M_{H}$ in the corresponding spacelike direction (as claimed in \cite{SeH2}).

Finally, we discuss the possibility of obtaining generalizations  of the PI (\ref{pi}). 
For spacetimes with electromagnetic fields the inequality $M \ge |Q|$, where $Q$ is the electric charge  
\cite{GHHP,MH}, has been shown (also in the presence of apparent horizons) whenever the norm of the charge 
4-current is not larger than  the norm of the matter 4-current. Moreover, both for the case $k_{ij} = 0$ 
as well as for general horizons in spherically symmetric spacetimes, it is known that 
$ M \ge \sqrt{A/16\pi} +  Q^2\sqrt{\pi/A}$ provided all charges are inside
${\cal H}$ \cite{MM,PJ2}.
Under the same requirements, the same generalized PI can be shown for general 
non-spherical horizons by applying the same arguments as in \cite{PJ2} to our final expression (\ref{Lm}).  
One could also try to incorporate the linear momentum into the inequality. 
It might be possible to get (\ref{gpi}) (or something similar) by extracting 
the momentum out of (\ref{Lm}) or, perhaps better, by looking for an alternative 
energy functional which directly gives $M^2 - P^2$ at infinity. 

\bigskip

{\bf Acknowledgements.}  E.M. was  supported in part by
the KBN grant 2 PO3B 010 16, and W.S. was supported by 
the Austrian  FWF project P14621-Mat. W.S. is grateful to Helmuth Urbantke for helpful discussions.

\end{document}